\documentclass[11pt,a4paper]{article}
\usepackage[english]{babel}
\usepackage[isolatin]{inputenc} 

\usepackage{etaremune}
\usepackage{amsfonts,amssymb,amsmath,bbm}
\usepackage{fontenc,indentfirst, delarray}
\usepackage{graphicx,graphics,xcolor,epstopdf,epsf}
\DeclareGraphicsExtensions{eps,ps,p.gz,pdf}
\usepackage{pdfsync}
\usepackage[pdfstartview=FitH]{hyperref}
\usepackage{setspace}

\usepackage{footmisc}

\makeatletter
\renewcommand{\section}{\@startsection {section}{1}{\z@}%
             {-3.5ex \@plus -1ex \@minus -.2ex}%
             {2.3ex \@plus.2ex}%
             {\normalfont\normalsize\sffamily\bfseries}}
\renewcommand{\subsection}{\@startsection {subsection}{1}{\z@}%
             {-3.5ex \@plus -1ex \@minus -.2ex}%
             {2.3ex \@plus.2ex}%
             {\normalfont\normalsize\sffamily\emph}}

\@addtoreset{equation}{section}

 \makeatother
\definecolor{bleuvert}{rgb}{.1,.5,.4}
\definecolor{light-gray}{gray}{0.95}
\definecolor{gray}{gray}{0.75}
\definecolor{violet}{rgb}{0.4,0.,0.3}
\definecolor{jaune}{rgb}{0.8,0.6,0.1}
\definecolor{cvert}{rgb}{0.8,0.6,0.5}

\newtheorem{thm}{Theorem}[section]

\newtheorem{prop}[thm]{Proposition}

\newenvironment{preuve}{{\emph{Proof.}}}{\hfill$\blacksquare$}

\newcommand{\norm}[1]{\left\lVert#1\right\rVert}
\newcommand{\abs}[1]{\lvert#1\rvert}

\newcommand{\grad}[1]{\text{grad}\,#1}

\newcommand{\ket}[1]{\lvert#1\rangle}
\newcommand{\bra}[1]{\langle#1\lvert}

\def\begf{\begin{frame}}
\def\enf{\end{frame}}

\def\begz{\begin{itemize}}

\def\endz{\end{itemize}}

\def\lp{\left(} 
\def\rp{\right)} 
\def\dm{\lp\begin{array}}	
\def\fm{\end{array}\rp}
\def\m2{M_2 \lp \cc \rp}
\def\m3{M_3 \lp \cc \rp}

\def\ds{\partial\!\!\!\slash}

\def\cc{{\mathbb{C}}}
\def\C{{\mathbb{C}}}

\def\ii{{\mathbb{I}}}

\def\M{{\mathcal M}}		
\def\aa{{\mathcal A}}

\def\A{{\mathcal A}}			
\def\bb{{\mathcal B}}

\def\hh{{\mathcal H}}

\def\pp{{\mathcal P}}

\def\X{{\mathcal X}}

\def\ss{{\mathcal S}}

\def\lda{\text{Lip}_D(\A)}

\def\xo0{\omega^0_x}
\def\yo0{\omega^0_y}

\def\xo0{x_\omega^0}
\def\yo0{y_\omega^0}

\def\pa{{\cal P}(\aa)}
\def\sa{{\cal S}(\aa)}

\def\fm{\Phi(x^\mu)}

\def\dm{\partial_\mu}

\def\X{{\cal X}}
\def\NN{{\cal N}}
\def\dmm{\left(\begin{array}}
\def\fmm{\end{array}\right)}

\newcommand{\HH}{\mathcal{H}}
\newcommand{\de}{\mathrm{d}}
\newcommand{\abso}[1]{|#1|}

\begin{document}
\title{\vspace{-2.5truecm}Towards a Monge-Kantorovich metric\\ in noncommutative geometry}
\author{Pierre Martinetti{\footnote{{\small martinetti.pierre@gmail.com}}}\\
{\footnotesize CMTP \& Dipartimento di Matematica,~Universit\`a di Roma Tor Vergata,
 I-00133}\\
{\footnotesize Universit\`a di
  Napoli Federico II, I-00185}}
\date{Proceeding of the international conference for the Centenary of Kantorovich,\\
    \emph{Monge-Kantorovich optimal transportation problem, transport metric
  \\ and their applications},  St-Petersburg, June 2012.}
\maketitle

\abstract{We investigate whether the identification between Connes'
  spectral distance in noncommutative geometry  and the Monge-Kantorovich distance of order $1$ in the theory of optimal
  transport - that has been pointed out by Rieffel in the commutative
  case - still
  makes sense in a noncommutative framework. To this aim, given a
  spectral triple $(\A, \hh, D)$ with noncommutative $\A$, we
  introduce a ``Monge-Kantorovich''-like distance $W_D$ on the space
  of states of $\A$, taking as a cost function the spectral distance $d_D$
  between pure states. We show in full generality that $d_D\leq W_D$, and exhibit several examples where the equality actually
  holds true, in particular on the unit two-ball viewed as the state
  space of $M_2(\C)$. We also discuss $W_D$ in a two-sheet model
  (product of a manifold by $\C^2$), pointing towards a possible
  interpretation of the Higgs field as a cost function that does not
  vanish on the diagonal.}

\section{Introduction}

In \cite{Connes:1989fk} Connes noticed that the geodesic distance on a compact Riemannian
mani\-fold $\M$ (connected and without boundary) can be retrieved in purely
algebraic terms, from the
knowledge of both the algebra $C^\infty(\M)$ of smooth functions on $\M$
and the signature (or Hodge-Dirac operator) $d + d^\dag$, where $d$ is the exterior
derivative. Explicitly, one has
\begin{equation}
  \label{eq:1}
  d_{\text{geo}}(x, y) = \sup_{f\in C^\infty(\M)} \left\{ \abs{\delta_x(f) - \delta_y(f)},\; \norm{[d+d^\dag,\pi(f)]}\leq 1\right\}
\vspace{-.35truecm}\end{equation}
where
\begin{itemize}\item [-]$\pi$ denotes the representation of the commutative algebra $C^\infty(\M)$ by multiplication on
  the Hilbert space $\Omega^\bullet(\M)$ of square integrable differential forms on
  $\M$;
\item[-]the norm of the commutator is the operator norm on $\bb(\Omega^\bullet(\M))$
  (bounded operators on $\Omega^\bullet(\M)$): 
\begin{equation}
\norm{A}= \sup_{\norm{\psi}_{\Omega}=1}\norm{A\psi}_{\Omega}
\label{eq:17}
\end{equation}
 for any $A\in\bb(\Omega^\bullet(\M))$, with $\norm{\cdot}_{\Omega}$ the Hilbert space
 norm of $\Omega^\bullet(\M)$;
\item[-] $\delta_x: f\to
  f(x)$ is the evaluation at $x\in\M$.
\end{itemize}

Evaluations are nothing but the pure states of the $C^*$-closure $C(\M)$
of $C^\infty(\M)$. Recall that a state of a $C^*$-algebra $\A$ is a
positive linear form on $\A$ with norm $1$. The space of states,
denoted $\sa$, is convex and its extremal points are called \emph{pure
  states}. 
By Gelfand theorem, any commutative $C^*$-algebra $\A$ is isomorphic
to  the algebra of functions vanishing at infinity on its pure state
space $\pa$ and - conversely - any locally compact topological space
$\X$ is homeomorphic to
the pure state space of $C_0(\X)$,
\begin{equation}
  \label{eq:8}
 \A\simeq C_0(\pa) \quad   \X \simeq {\mathcal P}(C_0(\X).
\end{equation}
In modern terms, the category of commutative $C^*$-algebras is
(anti)-isomorphic to the category of locally compact topological
spaces. The compact case corresponds to unital algebras.

With Gelfand theorem in minds, it is natural to extend (\ref{eq:1}) to non-pure states $\varphi,
\tilde\varphi\in{\cal S}(C(\M))$, defining
\begin{equation}
  \label{eq:1bis}
  d_{d+d^\dag}(\varphi, \tilde\varphi) \doteq \sup_{f\in C^\infty(\M)} \left\{ \abs{\varphi(f) - \tilde\varphi(f)},\; \norm{[d+d^\dag,\pi(f)]}\leq 1\right\}.
\end{equation}

Since the commutativity of the
algebra $C^\infty(\M)$ does not enter (\ref{eq:1}), another natural
extension is to the noncommutative setting. Namely, given a noncommutative algebra $\A$ acting by $\pi$ on some
Hilbert space $\hh$, together with an operator $D$ on $\hh$ such that $[D,
\pi(a)]$ is bounded for any $a\in\A$, one defines \cite{Connes:1994kx} for any $\varphi,
\tilde\varphi\in\sa$ 
\begin{equation}
  \label{eq:1ter}
  d_D(\varphi, \tilde\varphi) \doteq \sup_{a\in\bb_D(\A)} \abs{\varphi(a) - \tilde\varphi(a)},
\end{equation}
where 
\begin{equation}
  \label{eq:7}
  \bb_D(\A)\doteq \left\{ a\in \A, \norm{[D,\pi(a)]}\leq 1\right\}
\end{equation}
denotes the \emph{$D$-Lipschitz ball of $\A$}.  It is easy to check that $d_D$ satisfies all the properties of a
 distance on $\sa$ (see e.g. \cite [p.35]{Martinetti:2001fk}), except it may be infinite. Following the terminology of
 \cite{Martinetti:2008hl},  we call it the
\emph{spectral distance}{\footnote{Because it is a distance associated
    with a spectral triple, cf section~\ref{spectriple}}, but depending on the
authors it may be called Connes or the noncommutative
distance. Notice that - as in the commutative case - when $\A$ is
not a $C^*$-algebra we consider the states of its $C^*$-closure in the operator norm coming from the
representation $\pi$. 

Therefore the spectral distance (\ref{eq:1ter}) appears as a generalization to the
noncommutative setting of the Riemannian geodesic distance. The latter
is retrieved between pure states in the  commutative case
\begin{equation}
d_{d+d^\dag}(\delta_x, \delta_y) = d_{\text{geo}}(x,y).\label{eq:13}
\end{equation}
For non-pure states (still in the commutative case), Rieffel
seems to have been the first to notice in \cite{Rieffel:1999ec} that (\ref{eq:1bis})
was nothing but Kantorovich's dual formulation of the minimal
transport between probability measures, with cost function the geodesic distance.
Indeed, the set of states of $C(M)\ni\varphi$ is in $1$-to-$1$ correspondence with
the set of probability measures $\text{Prob}(\M)\ni\mu$ on $\M$,
\begin{equation}
  \label{eq:10}
  \varphi(f) = \int_\M f \,\de\mu,
\end{equation}
and it is not difficult to check (as recalled in section \ref{commutative}) that
\begin{equation}
  \label{eq:11}
  d_{d+d^\dag}(\varphi, \tilde\varphi) = W(\mu, \tilde\mu)
\end{equation}
where $W$ denotes the Monge-Kantorovich (or Wasserstein) distance of
order one with cost $d_{\text{geo}}$.{\footnote{In this contribution we are only interested in the
    Monge-Kantorovich distance of order $1$ with cost the geodesic
    distance. From now on we simply call it the Monge-Kantorovich distance..}} The same result holds
on a locally compact manifold, as soon as it is complete \cite{dAndrea:2009xr}. 
\newline

In this contribution, we investigate how the identification of the
spectral distance with the Monge-Kantorovich metric
could still make sense
in a noncommutative context. Namely, given~a~non\-commutative algebra
$\A$ acting on some Hilbert $\hh$ together with an operator $D$ such
that $[D,\pi(a)]$ is bounded for any $a\in\A$, is there some ``optimal
transport in noncommutative geometry'' such that the associated
Monge-Kantorovich distance coincides with the spectral distance~(\ref{eq:1ter})~?
We provide a tentative answer, introducing on the state space $\sa$ a
new distance $W_D$, obtained by taking as a cost function the spectral
distance $d_D$ on the pure state space $\pa$. The main properties of
this ``Monge-Kantorovich''-like distance
are worked out in proposition \ref{propmkncg}: it is shown in full
generality that
$d_D\leq W_D$ on $\sa$, with equality on $\pa$ as well
as between any  non-pure states obtained as convex linear combinations of the same
two pure states. In particular, this allows to show that $d_D = W_D$ on
the two-ball, viewed as the space of states of $\A=M_2(\C)$.

We begin with the commutative case (section \ref{commutative}),
recalling in proposition \ref{propcomm} why $d_{d+d^\dag}$
coincides with the Monge-Kantorovich distance $W$. Then we investigate the
noncommutative case and introduce the new distance $W_D$ in section
\ref{sectionncg}.
Section \ref{examples} deals with examples.

\section{The commutative case}
\label{commutative}

\subsection{Monge-Kantorovich and spectral distance}
Recall that given two probability measures $\mu$, $\tilde\mu$ on a
metric space $(\X, d)$ (non-necessarily compact), the Monge-Kantorovich distance is
\begin{equation}
  \label{eq:12}
  W(\mu,\tilde\mu) = \inf_{\pi}\int_{\X\times\X}\!\!\!\!\!\!\!\!\!\!\!\!\de\pi\quad d(x,y)
\end{equation}
where the infimum is on all the measures on $\X\times\X$ whose
marginals are $\mu$ and $\tilde\mu$.
 In his seminal work \cite{Kan42,KR58}, Kantorovich
showed there exists a dual formulation, 
\begin{equation}
  \label{eq:12bis}
 W(\mu, \tilde\mu) = \sup_{\norm{f}_{\mathrm{Lip}}\leq 1, \,f\in L ^1(\mu_1)\cap L ^1(\mu_2) }\left(\int_\X f\de\mu - \int_\X f\de\tilde\mu\right)
\end{equation}
for any pair of probability measures on $\X$ such that the right-hand side in the above expression is finite.
The supremum is on all real $\mu,\tilde\mu$-integrable real functions
$f$ that are 1-Lipschitz, that is to say 
\begin{equation}
  \abso{f(x) -f(y)} \leq d(x,y) \quad\forall\; x,y \in
  \X.
\label{eq:25}
  \end{equation}

 Take now $(\X, d)$ a locally compact Riemannian manifold $(\M,d_{\text{geo}})$ and consider the spectral distance (\ref{eq:1bis}). The formula
 is the same as in the compact case, except that we want the algebra
 to be  represented by bounded operators. So instead of
 $C^\infty(\M)$  we  look for the supremum on the algebra $C^\infty_0(\M)$
 of smooth functions vanishing at infinity.
Let $\varphi, \tilde\varphi$ be two states of $C_0(\M)$
 defined by probability measures $\mu, \tilde\mu$ via formula
 (\ref{eq:10}).  That the Monge-Kantorovich distance $W(\mu,
 \tilde\mu)$ equals the spectral distance $d_{d+d^\dag}(\varphi,
 \tilde\varphi)$ follows from the three
 well known points:
 \begin{itemize}

\item[-]  the supremum in (\ref{eq:1ter}) can
   be equivalently searched on selfadjoint elements
   \cite{Iochum:2001fv}. In the commutative case, this means we assume $f\in C_0^\infty(\M)$ is real.

  \item[-] for real functions, the norm of the commutator $[d +d^\dagger,
   \pi(f)]$,  as an operator on $\Omega^\bullet(\M)$, is precisely the
   Lipschitz norm of $f$ (see \cite{Connes:1992bc} and also section \ref{secspin}):
   \begin{equation}
   \norm{f}_{\text{Lip}} = \norm{[d+d^\dag,\pi(f)]}.
\label{eq:16}
    \end{equation}

 \item[-] the supremum on $1$-Lipschitz smooth functions vanishing at
   infinity in the spectral distance formula  is the same as the
   supremum on $1$-Lipschitz continuous functions non-necessarily
   vanishing at infinity in Monge-Kantorovich formula  (for details cf e.g. \cite[\S 2.2]{dAndrea:2009xr}).
 \end{itemize}
   Notice that for the last point to be true, it is important that $\M$ be
complete. Under this condition one obtains 
\begin{prop}\cite{Rieffel:1999ec,dAndrea:2009xr}
\label{propcomm}
  On a (connected, without boundary) complete Riemannian manifold
  $\M$, for any state $\varphi, \tilde\varphi\in S(C_0(\M))$ one has
  \begin{equation}
    \label{eq:14}
    d_{d+d^\dag}(\varphi, \tilde\varphi) = W(\mu, \tilde\mu).
  \end{equation}
\end{prop}

\subsection{On the importance of being complete}
 
It is not known to the author whether Kantorovich duality holds for non-complete
manifolds (in the literature the completeness condition seems to be
always assumed). In any case, (\ref{eq:12bis}) still makes sense as a
definition of the Monge-Kantorovich distance for non-complete
manifolds. The importance of the completeness
condition is illustrated by simple examples, taken from \cite{dAndrea:2009xr}.

Let $\M$ be a compact manifold and $\NN= \M\smallsetminus \left\{  x_0 \right\}.$
For example $\M = S^1 = [0,1]$ and $\NN=(0,1)$. The
Monge-Kantorovich distance on $S^1$ is
\begin{equation}
W_{\M}(x,y) =
\min\{|x-y|,1-|x-y|\},\label{eq:15}
\end{equation}
which differs from $W_{\NN}(x,y) = |x-y|$.
On the contrary, for $\M=S^2$ and $\NN=S^2\smallsetminus\left\{ x_0\right\}$, one has $W_\M = W_\NN$.
Removing a point from a complete compact mani\-fold may change or not
 the Monge-Kantorovich distance. 

On the contrary, removing a point does not modify the spectral distance, in the sense that 
\begin{align*}
  d_{d+d^\dag}^{\M}(\varphi_1,\varphi_2) &= \sup_{f\in C^\infty(\M)} \big\{
  \abso{(\varphi_1 - \varphi_2)(f)};\, ||f||_{\mathrm{Lip}}\leq
  1\big\} & \\ &= \sup_{f\in
    C^\infty(\M),f(x_0)=0} \big\{ \abso{(\varphi_1 - \varphi_2)(f)};\, ||f||_{\mathrm{Lip}}\leq 1\big\}  \\ &= \sup_{f\in
    C^\infty_0(\NN)} \big\{ \abso{(\varphi_1 - \varphi_2)(f)};\, ||f||_{\mathrm{Lip}}\leq 1\big\} = d_{d+d^\dag}^{\NN}(\varphi_1,\varphi_2).
\end{align*}
Here we noticed that because $ C^\infty(\M)$ has a
unit ${\bf 1}$, if $f$ attains the supremum then so does $f-
f(x_0){\bf 1}$ (the argument is still valid if the supremum is not
attained, by considering a sequence of element in the Lipschitz ball
tending to the infimum). 

To summarize, the spectral and the
Monge-Kantorovich  distances are equal on the incomplete manifold
$S^2\smallsetminus\left\{x_0\right\}$, but are not equal on $(0,1)$.

\subsection{Spin, Laplacian and the Lipschitz ball}
\label{secspin}

There exist alternative definitions of the Lipschitz ball
(\ref{eq:7}). Instead of the signature operator $d +d^\dag$, one can use as well the Dirac (or
Atiyah) operator
\begin{equation}
\ds=-i\sum_{\mu=1}^{\text{dim} \M}\gamma^\mu \partial_\mu.
\label{eq:18}
\end{equation} 
Recall that the $\gamma^\mu$'s are selfadjoint
matrices of dimension $M~\doteq~2^{E(\frac m2)}$, $m\doteq \text{dim}
\,\M$, spanning an irreducible representation of the
Clifford algebra. They satisfy
\begin{equation}
\gamma^\mu\gamma^\nu +
\gamma^\nu\gamma^\mu = 2g^{\mu\nu}\ii_M.\label{eq:19}
\end{equation} 
With $\pi_1$ the multiplicative representation of $C_0(\M)$ on the Hilbert space $\hh_1$
of square integrable spinors on $\M$,
\begin{equation}
\left(\pi_1(f)\psi\right)(x) = f(x)\psi(x) \quad \forall \psi\in\hh_1,\,x\in\M,
\label{eq:23}
\end{equation}
one easily checks that $[\ds, \pi_1(f)]$ acts as multiplication by
$\sum_{\mu}\gamma^\mu\partial_\mu f$, since by the Leibniz rule
\begin{equation}
  \label{eq:22}
  [\ds, \pi_1(f)]\psi = \sum_{\mu}\gamma^\mu\partial_\mu f \psi - f\gamma^\mu\partial_\mu \psi =  \left(\sum_{\mu}\gamma^\mu\partial_\mu f\right)\psi.
\end{equation}
Hence for real functions $f$, using the property of the $C^*$-norm and
Einstein summation on repeated indices, one gets 
\begin{align}
  \label{eq:24}
  \norm{[\ds, \pi_1(f)]}^2 &= \norm{(\gamma^\mu\partial_\mu
    f)^*\gamma^\nu\partial_\nu f} = 
\norm{\gamma^\mu\gamma^\nu\partial_\mu f\partial_\nu f}
= \frac 12\norm{(\gamma^\mu\gamma^\nu +\gamma^\nu\gamma^\mu) \partial_\mu
    f\partial_\nu f} \\& = \norm{g^{\mu\nu}\partial_\mu f \partial_\nu f}
  = \norm{\grad f} = \norm{f}_{\text{Lip}}.
\end{align}

The Lipschitz norm of $f$ can also be
retrieved from the Laplacian $\Delta$ (see e.g. \cite[\S
2.2]{dAndrea:2009xr} for details)
\begin{equation}
  \label{eq:20}
  \norm{f}_{\text{Lip}} = \tfrac{1}{2} \norm{[[\Delta ,\pi_2(f)] , \pi_2(f)]}
\end{equation}
where $\pi_2$ denotes the representation of $C_0(\M)$ on the Hilbert space
of square integrable functions on $\M$.
In the noncommutative setting, one could be tempted to define the Lipschitz ball
with a bi-commutator similar to (\ref{eq:20}) instead of (\ref{eq:7}).
However, it is easier to generalize to the noncommutative
case a first order differential operator than a second order one, which
justifies the choice of (\ref{eq:7}).

As proposed by Rieffel, one could even work  with the unit ball
\begin{equation}
\bb_L\doteq \left\{a\in\A, \,L(a)\leq
  1\right\}
\label{eq:21}
\end{equation}
with $L$ a seminorm not necessarily coming from the
commutator with an operator. In this contribution however, having Connes' reconstruction
theorem in minds (see the next section) we will
stick to the definition (\ref{eq:7}), and view $D$ as a noncommutative
generalization of the Dirac operator. 

\section{A Monge-Kantorovich metric in noncommutative geometry}
\label{sectionncg}
\subsection{Spectral triple}
\label{spectriple}

To extend formula  (\ref{eq:1bis}) to the noncommutative setting (\ref{eq:1ter}),
 the starting point is to choose a suitable
algebra $\A$, a suitable representation $\pi$ on some Hilbert space $\HH$, and
a suitable operator $D$. For (\ref{eq:1ter}) to make sense as a
distance,  one needs that $[D,\pi(a)]$ be in $\bb(\hh)$ for any
$a\in\A$, or at least for a dense subset of $\A$; otherwise one may
have $d_D(\varphi, \tilde\varphi)= 0$ although $\varphi\neq
\tilde\varphi$. Following Connes \cite{Connes:1996fu} one further asks that
\begin{itemize}
  \item[0.]
$\pi(a)[D - \lambda\ii]$ is compact for any $a\in\A$ and $\lambda$ in the
  resolvent set of $D$.
\end{itemize}
When the algebra is unital, this simply means that $D$
has compact resolvent.
 A triplet $(\A, \hh, D)$ satisfying the conditions above is called a \emph{spectral triple}. 

Although for our purposes we do not need the full machinery of
noncommutative geometry, it is interesting to recall the general
context. By imposing five extra-conditions{\footnote{They are quite technical and we do
not need them here. Let us simply mention that can be viewed as an algebraic
translation of the following properties of a manifold: i. the
dimension, ii. the signature operator being
a first order differential operator iii. the smoothness of
the coordinates, iv. orientability, v. existence of the tangent bundle.}} Connes is able to extend Gelfand duality beyond
topology \cite{connesreconstruct}: if $(\A,\hh, D)$ is a spectral
triple satisfying i-v with $\A$
unital \& commutative, then there exists a compact (connected, without
boundary) Riemannian manifold $\M$ such
that $\A\simeq C^\infty(\M)$. Conversely, to any such $\M$ is associated the spectral triple $(C^\infty(\M),
\Omega^\bullet(\M), d+d^\dag)$ which satisfies i-v. With two more
conditions (vi real structure, vii Poincaré duality), the reconstruction theorem extends to
spin manifolds.

A noncommutative
geometry is intended as a geometrical object whose set of functions defined on it
is a noncommutative algebra. As such it is not a usual space (otherwise its algebra of functions would be
commutative, by Gelfand theorem), so it requires new mathematical
tools to be investigated. Spectral triples provide such tools: first by
formulating in purely algebraic terms all the aspects of Riemannian
geometry (Connes reconstruction theorem), second by giving them a
sense in the noncommutative context (properties i-vii still
makes sense for noncommutative $\A$). 
\begin{eqnarray*}
\text{commutative spectral triple} &\rightarrow& \text{noncommutative
  spectral triple}\\
\updownarrow & & \downarrow \\
\text{Riemannian geometry} & & \text{non-commutative geometry}
\end{eqnarray*}

Specifically, the formula (\ref{eq:1ter}) of the spectral distance
is a way to export to the noncommutative setting the usual notion of
Riemannian geodesic distance. Notice the change of
point of view: the distance is no longer the infimum of a
geometrical object (i.e. the length of the
paths between points), but the supremum of an algebraic quantity
(the difference of the valuations of two states). This is interesting
for physics, for it provides a notion of distance no longer
based on objects ill defined in a quantum context:  
Heisenberg uncertainty principle makes the notions of points and path
between points highly problematic. 

A natural question is whether one
looses any trace of the distance-as-an-infimum  by passing to the
noncommutative side. More specifically, is there some
``noncommutative Kantorovich duality'' allowing to view the spectral
distance as the minimization of some ``noncommutative cost'' ? 
\begin{align*}
\text{\underline{distance as a supremum}:} & \hspace{1truecm} d_{d+d^\dag}  \text{ commutative case}
&\rightarrow&\quad d_D \text{ noncommutative case}\\
&\hspace{1truecm} \uparrow & & \quad\quad\;\lvert \\
 \text{Kantorovich duality: } &\hspace{1truecm} d_{d+d^\dag} = W& &\quad d_D = W_D  ? \\
&\hspace{1truecm} \downarrow & & \quad\quad\downarrow\\
\text{\underline{distance as an infimum}:}&\hspace{1truecm}  W \text{
  with cost } d_{\text{geo}}& & \quad\text{noncommutative cost ?}\\
\end{align*}
The (very preliminary) elements of answer we give in the next section comes from the following
observation: in the commutative case, the cost function is retrieved as
the Monge-Kantorovich distance between pure states of $C_0(\M)$. So in
the noncommutative case, if
the spectral distance were to coincide with some
``Monge-Kantorovich''-like distance $W_D$ on $\sa$,
then the associated cost should be the spectral distance on the pure
state space $\pa$. 

\subsection{Optimal transport on the pure state space}
\label{optpurestate}

Let $(\A, \hh, D)$ be a spectral triple. We aim at defining a
``Monge-Kantorovich''-like distance $W_D$ on the state space $\sa$, taking as a cost function the
spectral distance $d_D$ on
the pure state space $\pa$. A first idea is to mimic formula
(\ref{eq:12}) with $\X = \pa$, that is
\begin{equation}
  \label{eq:30}
W(\mu, \tilde\mu) = \inf_{\pi} \int_{\pa\times\pa} \hspace{-1.25truecm} \de\pi \quad d_D(\omega, \tilde\omega).
\end{equation}
For this to make sense as a distance on $\sa$, we should restrict to states
$\varphi\in\sa$ that are given by a probability
measures on $\pa$.  This is possible (at least) when $\A$ is separable and
unital: $\sa$ is then metrizable
\cite[p. 344]{Bratteli:1987fk} so that by Choquet theorem any state
$\varphi\in\sa$ is given by a probability measure
$\mu\in\text{Prob}(\pa)$. One should be careful however that the correspondence is not $1$ to $1$:
$\sa\to\text{Prob}(\pa)$ is injective, but two distinct probability
measures $\mu_1, \mu_2$ may yield the same state $\varphi$. This
is because $\A$ is \emph{not} an algebra of continuous functions on
$\pa$ (otherwise $\A$ would be commutative). We give an explicit
example of such a non-unique decomposition in section \ref{sphere}.

Thus $W_D$ is not a distance on  $\text{Prob}(\pa)$, but on a quotient of
it, precisely given by $\sa$. This forbids us to define $W_D$ by
formula (\ref{eq:30}).
A possibility is to consider the infimum 
\begin{equation}
\inf_{\mu, \tilde\mu} \; W(\mu, \tilde\mu)
\label{eq:33}
\end{equation}
on all the probability measures $\mu,
\tilde\mu\in\text{Prob}(\pa)$ such that 
\begin{equation}
  \label{eq:29}
  \varphi (a) =\int_{\pa} \omega(a)\, \de\mu, \quad   \tilde\varphi (a) =\int_{\pa} \omega(a)\, \de\tilde\mu.
\end{equation}
However it is not yet clear that (\ref{eq:33}) is a distance on $\sa$.

Here we explore another way, consisting in viewing $\A$ as an
``noncommutative algebra of functions'' on $\pa$,
\begin{equation}
  \label{eq:3}
  a(\omega) \doteq \omega(a) \quad \forall \omega\in\pa, \, a\in\A;
\end{equation}
and define the set of ``$d_D$-Lipschitz
noncommutative functions''   in analogy with (\ref{eq:25})
\begin{equation}
 \lda\doteq \left\{
 a\in\A \;\text{ such that }\; \abs{a(\omega_1) - a(\omega_2)} \leq d_D(\omega_1, \omega_2) \quad
  \forall \,\omega_1, \omega_2\in\pa\right\}. 
\label{eq:5}
\end{equation}
By mimicking (\ref{eq:12bis}) we then defines for any $\varphi, \tilde\varphi\in \sa$
  \begin{equation}
    \label{eq:2}
    W_D(\varphi, \tilde\varphi) \doteq \sup_{a\in\lda}\abs{\varphi(a)
        -\tilde\varphi(a)}.
  \end{equation}
\begin{prop}
\label{propmkncg}
$W_D$ is a distance, possibly infinite, on $\sa$. Moreover for any $\varphi, \tilde\varphi\in\sa$,
  \begin{equation}
\label{inclusionn}
d_D(\varphi,\tilde\varphi) \leq W_D(\varphi,\tilde\varphi).
\end{equation}
The equation above is an equality on the set of convex linear combinations
$\varphi_\lambda \doteq \lambda\, \omega_1 + (1-\lambda) \,\omega_2$
of any two given pure states $\omega_1, \omega_2$: namely
for any $\lambda, \tilde\lambda\in [0,1]$ one has
\begin{equation}
  \label{eq:35}
  d_D(\varphi_\lambda, \varphi_{\tilde\lambda}) = \abs{\lambda -
    \tilde\lambda}\,d_D(\omega_1, \omega_2) = W_D(\varphi_\lambda, \varphi_{\tilde\lambda}).
\end{equation}
 \end{prop}
\begin{preuve} We first check that $W_D$ is a distance.
Symmetry in the
  exchange $\varphi\leftrightarrow \tilde\varphi$ is obvious, as well
  as $\varphi~=~\tilde\varphi\Rightarrow~W_D(\varphi,
  \tilde\varphi)~=~0$. The triangle inequality is immediate: for any
  $\varphi'\in\sa$ one has
\begin{align}
  \label{eq:6}
  W_D(\varphi, \tilde\varphi) &=\sup_{a\in\lda}\abs{\varphi(a) -\tilde\varphi(a)} \leq
  \sup_{a\in\lda}
\left( \abs{\varphi(a) - \varphi'(a)} + \abs{\varphi'(a)
    -\tilde\varphi(a)}\right)\\
\nonumber
&\leq 
\sup_{a\in\lda}\abs{\varphi(a) -\tilde\varphi'(a)} + 
\sup_{a\in\lda}\abs{\varphi'(a) -\tilde\varphi(a)} = W_D(\varphi, \varphi') + W_D(\varphi', \varphi).
\end{align}
A bit less immediate is $W_D(\varphi,
  \tilde\varphi) =0 \Rightarrow \varphi = \tilde\varphi$. To show
  this, let us first observe that 
\begin{equation}
\label{incl}
  \bb_D(\A)\subset \lda
\end{equation}
otherwise there would exist $a\in \bb_D(\A)$ and $\omega_1, \omega_2\in\pa$ such that
$\omega_1(a)-\omega_2(a) > d_D(\omega_1, \omega_2),$
which would contradict the definition of the spectral distance. 
Let us now assume $W_D(\varphi,
  \tilde\varphi) =0$. This means 
  $\varphi(a) = \tilde\varphi(a)$ for all $a\in\lda$. For $a\notin\lda$,~denote
  \begin{equation}
\lambda_a \doteq \inf_{\omega_1, \omega_2\in\pa}\,\frac{d_D(\omega_1, \omega_2)}{\omega_1(a) - \omega_2(a)}.
\label{eq:9}
  \end{equation}
One has
\begin{equation}
\frac 1{\norm{[D,\pi(a)]}}\leq \lambda_a < 1.
\label{eq:27}
\end{equation}
The r.h.s. inequality comes from $a\notin\lda$: there exists at least one pair
$\omega_1, \omega_2$ such that $\omega_1(a) - \omega_2(a)>
d_D(\omega_1,\omega_2)$. The l.h.s. inequality follows from the definition of the spectral
distance: any pair $\omega_1, \omega_2 \in \pa$ satisfies
\begin{equation}
d_D(\omega_1, \omega_2)
\geq \frac{\omega_1(a) -\omega_2(a)}{\norm{[D,\pi(a)]}},
\label{eq:28}
\end{equation}
which is well defined because $a\notin \bb_D(\A)$ by (\ref{incl}) so
that $\norm{[D,\pi(a)]}\neq 0$.
 In other terms $\lambda_a$ is finite and
non-zero, so that $\lambda_a a$ is in
$\lda$, meaning that $\varphi(\lambda_a a) -
\tilde\varphi(\lambda_a a)$ - hence $\varphi(a) -
\tilde\varphi(a)$ - vanish. So $\varphi = \tilde\varphi$ and $W_D$ is
a distance.

Eq. (\ref{inclusionn}) follows from (\ref{incl}): 
the supremum for $d_D$ is searched on a smaller set than
for $W_D$.

The first part of (\ref{eq:35}) comes from
\begin{equation}
  \label{eq:36}
  \varphi_\lambda(a) - \varphi_{\tilde\lambda}(a) = (\lambda -\tilde\lambda) \left(\omega_1(a) - \omega_2(a)\right),
\end{equation}
for this means
\begin{align}
 \label{eq:37}
  d_D(\varphi_\lambda, \varphi_{\tilde\lambda}) &= \sup_{a\in\bb_D(\A)}
  \abs{\lambda -\tilde\lambda} \left(\omega_1(a) - \omega_2(a)\right)
  = \abs{\lambda -\tilde\lambda} \sup_{a\in\bb_D(\A)}
  \left(\omega_1(a) - \omega_2(a)\right)\\ &= \abs{\lambda
    -\tilde\lambda} d_D(\omega_1, \omega_2).
\end{align}
The second part of (\ref{eq:35})  is obtained noticing that
(\ref{eq:36}) together with the definition of $\lda$ imply
\begin{equation}
  \label{eq:38}
  W_D(\varphi_\lambda, \varphi_{\tilde\lambda}) \leq \abs{\lambda -
    \tilde\lambda} \,d_D(\omega_1, \omega_2),
\end{equation}
that is $W_D(\varphi_\lambda,
\varphi_{\tilde\lambda}) \leq d_D(\varphi_\lambda,
\varphi_{\tilde\lambda})$ by (\ref{eq:37}), and the result by (\ref{inclusionn}). 
 \end{preuve}
\newline

The difference between $W_D$ and $d_D$ - if any - is entirely contai\-ned in the difference between
the $D$-Lipschitz ball (\ref{eq:7}) and
$\lda$ defined in (\ref{eq:5}). In the commutative case
$\A=C_0(\M)$, these two notions of Lipschitz function
coincide with the usual one, so that $d_D = W_D =W$. For the moment, we let as an open question whether in the
noncommutative case $d_D = W_D$ in
full generality.
In the next section we illustrate the equality (\ref{eq:35}) with various examples, including a
noncommutative one $\A=M_2(\C)$. 

\section{Examples}
\label{examples}

\subsection{A  two-point space}
\label{2points}
The spectral triple
\begin{equation}
\label{2point}
\aa = \cc^2, \quad \hh=\cc^2,\quad D=\left(\begin{array}{cc} 0 & m \\
\bar m & 0 \end{array}\right)
\end{equation}
where $m\in\cc$ and $a=(z_1, z_2)\in\A$ is represented by
\begin{equation}
\pi(z_1, z_2) = \left(\begin{array}{cc} z_1 & 0\\ 0&z_2\end{array}\right)
\end{equation}
describes a two-point space, for the algebra
$\C^2$ has only two pure states,
\begin{equation}
\label{statec2}
\delta_1 (z_1, z_2)\doteq z_1,\quad \delta_2 (z_1, z_2) \doteq z_2.
\end{equation}
Hence any non-pure state is of the form $\varphi_\lambda = \lambda \delta_1 +
(1-\lambda)\delta_2$ and by proposition  \ref{propmkncg} one knows
that that $d_D = W_D$ on the whole of $\sa$. 

It is easy to check explicitly that the two notions of Lipschitz
element coincide: one has
\begin{equation}
  \label{eq:31}
  \norm{[D,a]} = \abs{m(z_1-z_2)},
\end{equation}
hence by (\ref{eq:1ter}) 
\begin{equation}
d_D(\delta_1, \delta_2)= \frac{1}{\abs{m}}.
\end{equation}
Therefore $a=(z_1, z_2)\in\text{Lip}_D(\C^2)$ means $\abs{z_1-z_2}\leq
  \frac{1}{\abs{m}}$,  which is equivalent to $\norm{[D,a]}\leq
  1$. Hence $\text{Lip}_D(\C^2) =\bb_D(\C^2)$.

Although very  elementary (and commutative !), this example
illustrates interesting properties of the spectral distance: $\pa$ is a discrete space (hence
there is no notion of geodesic) but still the distance is
finite; the spectral distance on non-pure states is 
``Monge-Kantorovich''-like with cost the spectral distance on
pure states.

\subsection{The sphere}
\label{sphere}
Let us now come to the slightly more involved (and noncommutative)
example $\aa=M_2(\C)$.
Identifying a $2\times 2$ matrix with its natural representation on
$\C^2$, any unit vector $\xi\in\C^2$ defines a pure
state
\begin{equation}
\omega_\xi(a) = (\xi, a \,\xi) = \text{Tr} (s_\xi\, a)\qquad \forall a\in\A
\end{equation} where $(\cdot,\cdot)$ denotes the usual inner
product in $\C^2$ and $s_\xi\in M_2(\C)$ is the projection on $\xi$
(in Dirac notation $s_\xi = \ket{\xi}\bra{\xi}$). Two vectors
equal up to a phase define the same state, and any pure state is
obtained in this way. In other terms, the set of pure states of $M_2(\C)$ is the complex projective
plane $\cc P^1$, which is in $1$-to-$1$ correspondence with the
two-sphere  via
\begin{equation}
\xi=\left(\begin{array}{c}\xi_1\\ \xi_2\end{array}\right)  \in
\C P^1 \leftrightarrow {\bf x_\xi}= \left\{\begin{array}{ccc}
x_\xi &=& \text{Re}(\overline{\xi_1}\xi_2)\\
y_\xi &=& \text{Im}(\overline{\xi_1}\xi_2)\\
z_\xi &=& \abs{\xi_1}^2 - \abs{\xi_2}^2\end{array}\right. \in S^2.
\end{equation}

 A non-pure state $\varphi$ is determined by a probability
 distribution $\phi$ on $S^2$:
\begin{equation}
    \varphi(a) = \int_{S^2} \phi({\bf x}_\xi )\,\omega_\xi (a) \,\de
    {\bf x}_\xi 
\label{eq:39}
\end{equation}
for any $a\in M_2(\C)$, with $d {\bf x}_\xi $ the $SU(2)$ invariant
measure on $S^2$. However the correspondence between $\text{Prob}(S^2)$ and
$\ss(M_2(\C))$ is not $1$-to-$1$. One computes \cite[\S 4.3]{Cagnache:2009oe} that the density matrix
$s_\varphi$ such that
\begin{equation}
 \varphi(a) =\text{Tr}\left(s_\varphi \, a\right)\label{eq:43}
\end{equation}
actually depends on the barycenter of the probability measure $\phi$ only:
\begin{equation}
  s_\varphi= \left( \begin{array}{cc}
\frac{z_\phi+1}2 & \frac{x_\phi-iy_\phi}2 \\ \frac{x_\phi+ iy_\phi}2 & \frac{1-z_\phi}2\end{array}\right)
\end{equation}
where
\begin{equation}
{\bf x}_\phi = (x_\phi,  y_\phi,  z_\phi) \quad\text{ with }\quad
 x_\phi  \doteq \int_{S^2} \phi({\bf x}_\xi) \, x_\xi \,\de {\bf x}_\xi\label{eq:40}
 \end{equation}
and similar notation for $y_\phi$, $z_\phi$.

With the equivalence relation 
$\phi\sim \phi' \Longleftrightarrow {\bf x}_\phi = {\bf x}_{\phi'}$,
the state space
\begin{equation}
{\mathcal{S}}(\M_2(\C)) = {\mathcal{S}}(C(S^2))/\sim \;=  {\text{Prob}}(S^2)/\sim\label{eq:41}
\end{equation}
is homeomorphic to the Euclidean 2-ball:
\begin{equation}
   \ss(\M_2(\C))\ni  \varphi \overset{ }{\longrightarrow} {\bf x}_\phi \in {\mathcal B}^2 .
\label{eq:42}
\end{equation}
This means that any two states $\varphi, \varphi'$ are convex linear
combinations of the same two pure states. So by proposition \ref{propmkncg}
one has $d_D = W_D$ on the whole of $\ss(M_2(\C))$.
\newline

Depending on the choice of the representation and of the Dirac
operator, one deals with completely different cost functions: viewing $\A = M_2(\C)$ acting on $\hh = M_2(\C)$ as a truncation of the Moyal
plane \cite{Cagnache:2009oe}, one inherits a Dirac operator such that $d_D$ is finite on
$\pa$ (hence on $\sa$): \begin{equation}
d_D({\bf x}_\phi,{\bf x}_{\phi'})=
\sqrt{\frac{\theta}{2}}\times\begin{cases}
\cos\alpha\;\, d_{Ec} ({\bf x}_\phi, {\bf x}_{\phi'}) & \mathrm{when }\;\,\alpha\leq \frac{\pi}4,\\
\frac{1}{2\sin \alpha}\, d_{Ec} ({\bf x}_{\phi}, {\bf x}_{\phi'})  & \mathrm{when }\;\, \alpha\geq \frac{\pi}4,
\end{cases}\label{eq:44}
\end{equation}
where $d_{Ec} ({\bf x}_{\phi}, {\bf x}_{\phi'})=|{\bf x}_{\phi}-{\bf x}_{\phi'}|$ is the Euclidean distance
in the ball 
and $\alpha$ is the angle between the segment $[{\bf x}_{\phi},
{\bf x}_{\phi'}]$ and the horizontal plane $z_\xi =
\text{constant}$ (see  figure \ref{St-Pefigfig}).
\begin{figure}[ht*]
\begin{center}
\hspace{-0truecm}
\vspace{-1truecm}\mbox{\rotatebox{0}{\scalebox{.9}{\includegraphics{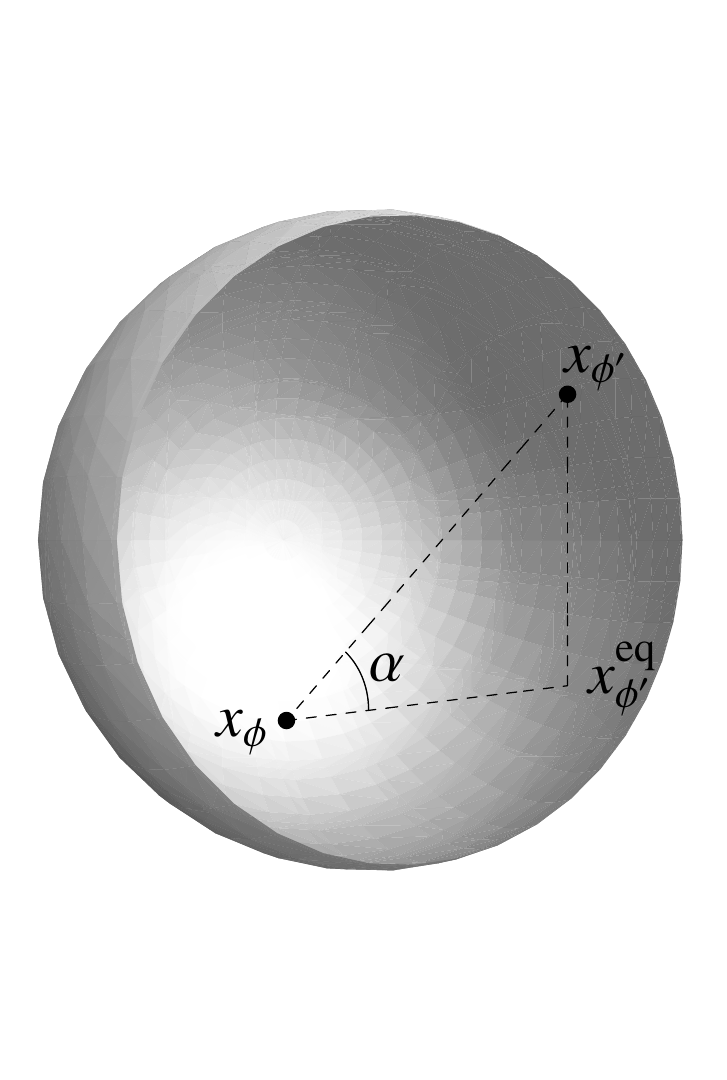}}}}
\end{center}
\caption{The vertical plane inside the unit ball that contains ${\bf
    x}_{\phi}$, ${\bf x}_{\phi'}$. We denote ${\bf x}_{\phi'}^{eq}$ the
  projection of ${\bf x}_{\phi'}$ in the ``equatorial'' plane $z_\phi
  = \text{cst}$}\label{St-Pefigfig} 
\end{figure} 

 On the contrary, making $\A = M_2(\C)$ act on $\hh = \C^2$, with $D$
 a $2$-by-$2$ matrix with distinct non-zero eigenvalues $D_1, D_2$, one obtains \cite{Iochum:2001fv}
\begin{equation}
d_D({\bf x}_\phi, {\bf x}_{\phi'}) = \left\{
  \begin{array}{ll}
\frac{2}{\abs{D_1-D_2}}\; d_{Ec} ({\bf x}_{\phi}, {\bf x}_{\phi'})&\text{ if } z_\phi = z_{\phi'},\\
 +\infty & \text{ if } z_\phi \neq z_{\phi'}.
 \end{array}
\right.
\end{equation}
Here the eigenvectors of $D$ - chosen as a basis of $\hh$ - are mapped to the north and south poles of $S^2$.
\subsection{Product of the continuum by the discrete}
 \label{product}
We summarize here the discussion developed in \cite[\S
4.2]{dAndrea:2009xr}. The
product of a compact manifold $\M$ by the spectral triple (\ref{2point})  is
the spectral triple $(\A', \hh', D')$ where \cite{Connes:1996fu}
\begin{equation}
\A' \doteq C^\infty(\M)\otimes \C^2,\; \hh' \doteq
  \Omega^\bullet(\M)\otimes\C^2,\; D' \doteq (d+d^\dagger)\otimes \ii_2 +
  \Gamma\otimes D\label{eq:45}
\end{equation}
with $\Gamma$ a graduation of  $\Omega^\bullet(\M)$. An element of
$\A'$ is a pair $a'=(f,g)$ of functions in $C^\infty(\M)$, and 
pure states of (the $C^*$-closure of) $\A'$ are pairs
\begin{equation}
x_i\doteq (\delta_x, \delta_i)
\label{eq:34}
\end{equation}
where $\delta_x\in\pp(C(\M))$ is the evaluation at $x\in\M$ while
$\delta_{i=1,2}$ is one of the two pure states of $\C^2$ defined in (\ref{statec2}).
  Thus $\pp(\A')$ appears as the disjoint union of
two copies of $\M$. Explicitly, the evaluation on an element of $\A'$ reads
\begin{equation}
 x_1(a') = f(x), \quad y_2(a') = g(y).
 \label{eq:32}
  \end{equation}
The spectral distance in this two-sheet model coincides with the
  geodesic distance $d'_{\text{geo}}$ in the manifold $\M'=\M\times[0,1]$ with Riemannian metric
  \begin{equation}
\left(\begin{array}{cc}
g_{\mu\nu} & 0 \\ 0 & \frac
1{\abs{m}}\end{array}\right)
\label{eq:48}
\end{equation}
where $g_{\mu\nu}$ is the Riemannian metric on $\M$. Namely one has \cite{Martinetti:2002ij}
\begin{equation}
  \label{eq:4}
  d_{D'}\left(x_1, y_2\right) = d'_{\text{geo}}\left( \left(x,0\right),
    \left(y, 1\right)\right). 
\end{equation}

Non-pure states of $\A'$ are given by pairs of measures $(\mu,\nu)$ on $\M$, normalized to 
$$\int_\M  \,\de\mu + \int_\M    \,\de\nu = 1,$$
whose evaluation on $a'= (f, g)$ is
\begin{equation}
\varphi (a) = \int_\M f \,\de\mu + \int_\M g \,\de\nu.
\label{eq:46}
\end{equation}
By proposition \ref{propmkncg} one has $d_{D'}\leq W_{D'}$ where
$W_{D'}$ is the Monge-Kantorovich distance on $\M \cup \M$ associated to the cost $d_{D'}$.
The equality holds for states
localized on the same copy:
$$\varphi=(0,\nu), \;\tilde\varphi=(0,\tilde\nu) \quad\text{ or }\quad
\varphi=(\mu,0),\; \tilde\varphi=(\tilde\mu,0),$$
since one then has
\begin{equation}
  \label{eq:47}
  d_{D'} (\varphi, \tilde\varphi)= d_{d+d^\dag}(\varphi,
  \tilde\varphi)= W(\varphi, \tilde\varphi) = W_{D'}(\varphi, \tilde\varphi). 
\end{equation}

For two states localized on distinct copies, the question is open. It
is interesting to notice that one may project back the problem on a
single copy of $\M$, using the
  cost function
$$c(x, y)  \doteq d_{D'}(x_1, y_2)  \doteq 
\sqrt{d_{\text{geo}}(x,y)^2 + \frac 1{\abs{m}^2}}$$
 defined on $\M$, rather than $d_{D'}$ defined on $\M \cup \M$.
The particularity of this single-sheet cost $c$ is that it does not
vanish on the diagonal, $c(x,x) = \frac 1{\abs{m}}\neq 0$.

This might yield interesting perspective in physics:  in the
description of the standard model of elementary particles in
noncommutative geometry \cite{Chamseddine:2007oz}, the recently
discovered Higgs field \cite{Collaboration:2012fk} appears as an
extra-component of the metric similar to $\frac 1{\abs{m}}$, except that it is no longer a constant but a
function on $\M$. From this perspective the Higgs field represents the cost to stay at the
    same point of space-time, but jumping from one copy to the
    other.

\section{Conclusion}
In this contribution we presented preliminary steps towards a definition of a
Monge-Kantorovich distance in noncommutative geometry. Given a
spectral triple $(\A, \hh, D)$, we introduced a new distance $W_D$
on the space of state $\sa$, which is the exact translation in the
noncommutative setting of Kantorovich dual formula, taking as a cost
function Connes spectral distance $d_D$ on the the pure state space
$\pa$. By construction, $W_D = d_D$ on $\pa$, and we showed that the
same is true for non-pure states given by convex linear combinations of the
same two pure states. Although very restrictive, this condition
applies to the interesting example $\A=M_2(\C)$, showing that $W_D =
d_D$ on a unit ball. 
At this point two questions remain open and
will be the object of future works:
\begin{itemize}
\item Is $W_D$ equal to $d_D$ on the whole of $\sa$ or only on part of
  it ?

\item Is there a dual formula to $d_D$ and/or $W_D$ as an infimum (a kind of reverse Kantorovich
  duality), for instance formula (\ref{eq:33}), or the Wasserstein
  distance in free probabilities introduced by Biane and Voiculescu
\cite{Biane:2001fkV} ?
\end{itemize}
These questions are not necessarily linked. If both answers turn out
to be positive, this would indicate that computing the spectral
distance is exactly a problem of optimal transport, and spectral
triples could be used as a factory of cost functions. 
\begin{center}
\rule{5cm}{.7pt}
\end{center}
\bibliographystyle{amsplain}
\bibliography{/Users/pierremartinetti/physique/articles/Bibdesk/biblio}

\end{document}